\documentclass{eas}
\usepackage{graphicx}
\pdfoutput=1
\usepackage{amsmath}
\usepackage{txfonts}
\usepackage[usenames,dvipsnames]{color}

\newcommand{\Msun}{\mathrm{M}_{\odot}}

\newcommand{\Mpmz}{M_{\mathrm{PMZ}}}

\newcommand{\Rsun}{\mathrm{R}_{\odot}}
\newcommand{\Hy}{\mathrm{H}}
\newcommand{\C}{\mathrm{C}}
\newcommand{\Cth}{^{13}\mathrm{C}}
\newcommand{\Ny}{\mathrm{N}}
\newcommand{\Ox}{\mathrm{O}}

\newcommand{\Fe}{\mathrm{Fe}}
\newcommand{\Sr}{\mathrm{Sr}}
\newcommand{\Yt}{\mathrm{Y}}
\newcommand{\Zr}{\mathrm{Zr}}
\newcommand{\Ba}{\mathrm{Ba}}
\newcommand{\La}{\mathrm{La}}
\newcommand{\Pb}{\mathrm{Pb}}

\begin{document}

\title{AGB nucleosynthesis \\at low metallicity: What can we learn from \\
Carbon- and s-elements-enhanced metal-poor stars}
\runningtitle{AGB nucleosynthesys at low Z: what can we learn from CEMP-$s$ stars}
\author{Carlo Abate}\address{Department of Astrophysics/IMAPP, Radboud University Nijmegen, P.O. Box 9010, 6500 GL Nijmegen, The Netherlands}
\author{Onno R. Pols$^1$}
\author{Robert G. Izzard}\address{Argelander Institut f\"ur Astronomie, Auf dem H\"ugel 71, D-53121 Bonn, Germany}
\author{Amanda I. Karakas}\address{Research School of Astronomy \& Astrophysics, Mount Stromlo Observatory, Weston Creek ACT 2611, Australia}
\begin{abstract}
CEMP-$s$ stars are very metal-poor stars with enhanced abundances of carbon and $s$-process elements. They form a significant proportion of the very metal-poor stars in the Galactic halo and are mostly observed in binary systems.
This suggests that the observed chemical anomalies are due to mass accretion in the past from an asymptotic giant branch (AGB) star. 
Because CEMP-$s$ stars have hardly evolved since their formation, the study of their observed abundances provides a way to probe our models of AGB nucleosynthesis at low metallicity. 
To this end we included in our binary evolution model the results of the latest models of AGB nucleosynthesis and we simulated a grid of 100,000 binary stars at metallicity Z=0.0001 in a wide range of initial masses and separations. We compared our modelled stars with a sample of 60 CEMP-$s$ stars from the SAGA database of metal-poor stars. For each observed CEMP-$s$ star of the sample we found the modelled star that reproduces best the observed abundances. The result of this comparison is that we are able to reproduce simultaneously the observed abundance of the elements affected by AGB nucleosynthesis (e.g. C, Mg, $s$-elements) for about $60\%$ of the stars in the sample.
\end{abstract}
\maketitle
\section{Carbon- and $s$-elements enhanced metal-poor stars.}
In the Galactic halo we observe a population of very metal-poor stars (VMP, $[\Fe/\Hy]\lesssim-2.0$) that are the relics of the early phase of star formation in our Galaxy. A fraction between $9\%$ and $25\%$ of the observed VMP stars are enriched in carbon (e.g. Marsteller \etal\ , \cite{Marsteller2005}; Frebel \etal\ , \cite{Frebel2006}; Lucatello \etal\ , \cite{Lucatello2006}). Among these carbon-enhanced metal-poor stars (CEMP, herein defined by $[\C/\Fe]\ge+1.0$), approximately $80\%$ are also enriched in heavy elements produced by slow ($s-$) and rapid ($r-$) neutron-capture processes (Aoki \etal\ , \cite{Aoki2007}). CEMP-$s$ stars are normally defined by the barium excess, namely $[\Ba/\Fe] \ge +1.0$. 

CEMP-$s$ stars are often found in binary systems: by analysing a sample of CEMP-$s$ stars Lucatello \etal\ (\cite{Lucatello2005b}) detected radial velocity variations in a fraction of stars statistically consistent with the hypothesis that all CEMP-$s$ stars are in binary systems. This evidence suggests that the origin of the peculiar abundances in CEMP-$s$ stars is due to mass transfer of carbon-rich material occurred in the past from an evolved primary star in the thermally-pulsing asymptotic giant branch (TP-AGB) phase. 
According to this interpretation, the study of CEMP-$s$ stars provides us with a powerful tool to probe AGB nucleosynthesis at low metallicity. Many CEMP-$s$ stars have hardly evolved since their formation and therefore the observed surface abundances are the fingerprint of the nucleosynthesis that occurred in their AGB companions. A correct understanding of CEMP-$s$ stars allows us to enlighten many different aspects of the AGB evolution at low metallicity that are poorly constrained, such as the physics of mixing, the mass-loss processes and the effects of binary interaction for example on the mass-accretion rate. 

\section{Nucleosynthesis during the AGB phase}
\label{det-model}
The AGB is the final stage in the life of stars with initial mass between about $0.8$ and $8\,\Msun$. An AGB star is characterised by two nuclear-burning shells, the innermost one burning helium (He) in unstable flashes above a degenerate carbon-oxygen core, and the other one burning hydrogen (H) below the convective envelope. When the He shell is active, the energy produced expands the outer layers and, as an effect of the reduced density, nuclear burning is estinguished in the H shell. At this stage the inner edge of the convective envelope can move inward (in mass) and mix to the surface the products of internal nucleosynthesis (Herwig, \cite{Herwig2005}), a process known as third dredge-up (TDU).

The $s$-process occurs in the intershell region between the two burning shells, where He is abundant and ($\alpha$, $n$) reactions can be efficiently activated to produce free neutrons, which are subsequently captured by the iron seeds to form $s$-elements. People usually define three peaks of $s$-elements that correspond to particularly stable atomic nuclei: the light-$s$ peak ($\Sr$, $\Yt$, $\Zr$), the heavy-$s$ peak ($\Ba$, $\La$), and the peak of lead ($\Pb$). Similarly to a system of three buckets of increasing sizes, each one containing the smaller ones and being contained in the larger, where to fill the largest bucket it is necessary to fill the smaller ones first, in an analogous way the $s$-process populates the three peaks progressively. When a small amount of free neutrons is available, the first $s$-elements to be produced are those in the light-$s$ peak. As soon as more free neutrons are available, the light-$s$ peak is filled and the heavy-$s$ elements are produced. When the density of free neutrons increases further also $^{82}\Pb$ is produced. 

The neutron source and the timescale for neutron production determine the density of free neutrons and ultimately the resulting $s$-process element distribution. 
For a star of mass up to $3\Msun$ the main source of free neutrons is the $^{13}\C(\alpha,\,n)^{16}\Ox$ reaction (Cameron, \cite{Cameron1955}). A sufficient abundance of $\Cth$ is produced in the intershell region when protons are captured by $^{12}\C$ to form a layer rich in $\Cth$ and $^{14}\Ny$
(Straniero, \cite{Straniero1995}). Hence, to allow the nucleosynthesis of $s$-elements, a mechanism is required that brings protons in the intershell region below the H-burning shell. During the TDU protons from the H-rich surface are mixed throughout the entire envelope. At the interface with the deepest extent of the convective envelope, some protons are mixed with the top layer of the intershell region. The details about the size in mass of this ``partial mixing zone'' (PMZ) and the physical process that creates the PMZ are still unknown, although several mechanisms have been proposed, such as convective overshoot, rotation and gravity waves (Herwig, \cite{Herwig2005}; Busso \etal\ \cite{Busso1999}). 

In the models computed by Karakas (\cite{Karakas2010}), protons are mixed into the intershell region by artificially adding the PMZ at the deepest extent of each TDU. For a given stellar mass the distribution of $s$-elements is determined by the mass of the PMZ, $\Mpmz$, because the larger the PMZ, the more free neutrons per iron seed are produced in the intershell. Karakas (\cite{Karakas2010}) and Lugaro \etal\ (\cite{Lugaro2012}) explore the effect of different $\Mpmz$ on the nucleosynthesis products of AGB stars of different initial mass.

\section{Results}
In this work we used the binary evolution model described by Izzard \etal\  (\cite{Izzard2004}; \cite{Izzard2006}; \cite{Izzard2009}) with the updates on the algorithm for the wind mass transfer described by Abate \etal\  (\cite{Abate2013}). In this model we implemented the intershell abundances derived by Karakas (\cite{Karakas2010}) and we simulated a grid of binary stars at metallicity $Z=10^{-4}$ ($[\Fe/\Hy]\approx-2.3$) where the initial parameters $M_1$, $M_2$, separation and $\Mpmz$ vary in the ranges $[0.9,\!6.0]\, \Msun$, $[0.4,\!0.9]\,\Msun$, $[10^2,\!10^4]\,\Rsun$ and $[0,\!0.004]\,\Msun$, respectively (details are provided in our forthcoming paper, Abate \etal\ , \cite{Abate2014}). Our aim is to reproduce the chemical abundances observed in a sample of 60 CEMP-$s$ stars selected from the SAGA database of metal-poor stellar abundances (Suda \etal\ \cite{Suda2008}; \cite{Suda2011}) and, in this way, to probe the models of AGB nucleosynthesis. 

For each CEMP-$s$ star of our sample we found the modelled star that matches best the observed chemical abundances by minimising the value of $\chi^2 = \Sigma_i (X_{i,\mathrm{obs}} - X_{i,\mathrm{mod}})^2/\sigma^2_{i,\mathrm{obs}}$, where $X_{i,\mathrm{obs}}$ and $X_{i,\mathrm{mod}}$ are respectively the observed and modelled abundance of the element $i$ and $\sigma_{i,\mathrm{obs}}$ is the observed error. To calculate $\chi^2$ we only used the elements that are affected by AGB nucleosynthesis, such as C, N, O, F, Na, Mg and neutron-capture elements. As an additional constraint, we imposed that the surface gravity of our modelled star reproduces the observed value within the error. 

\begin{figure}[ht!]
\includegraphics[width=0.56\textwidth]{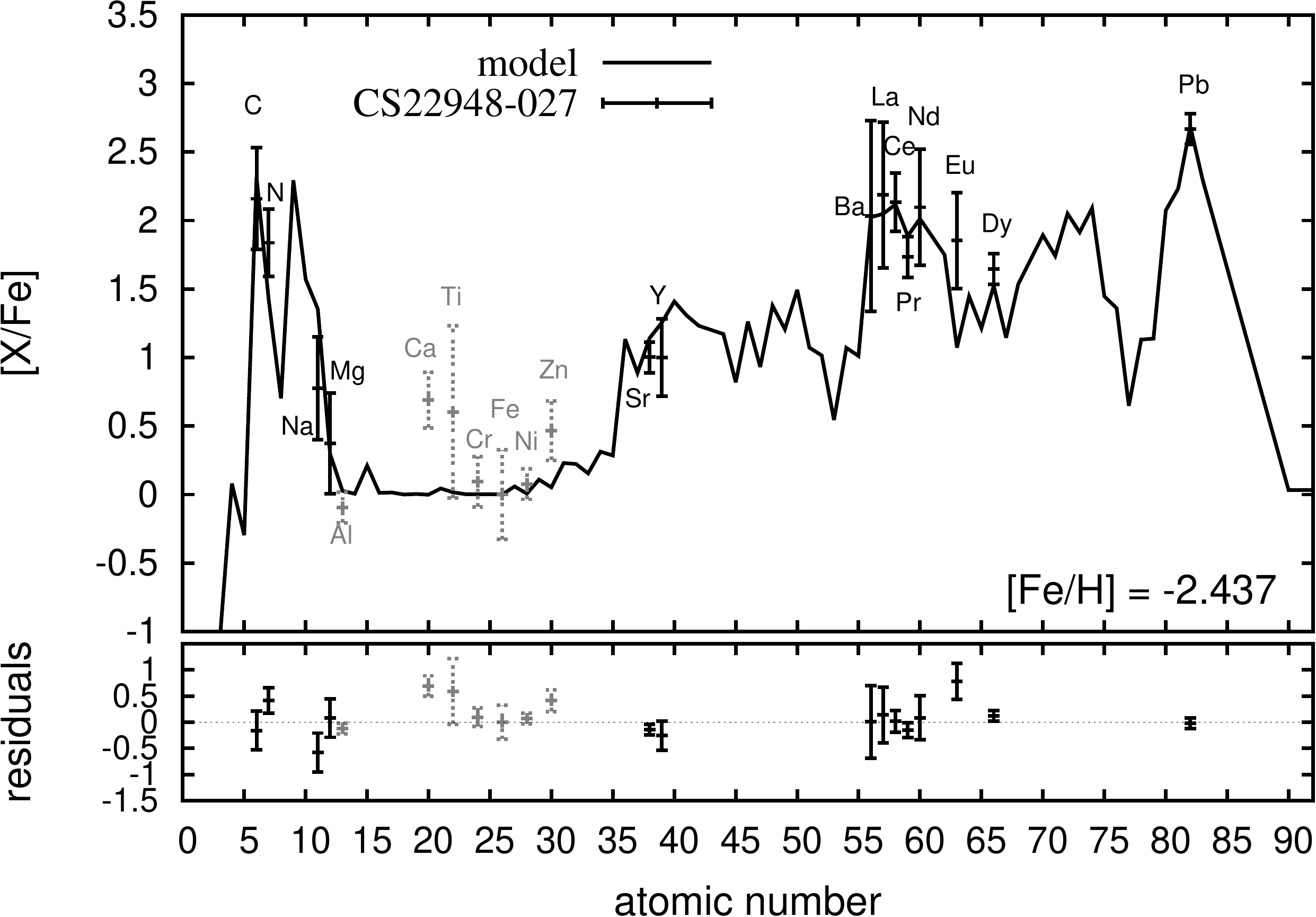}
\includegraphics[width=0.56\textwidth]{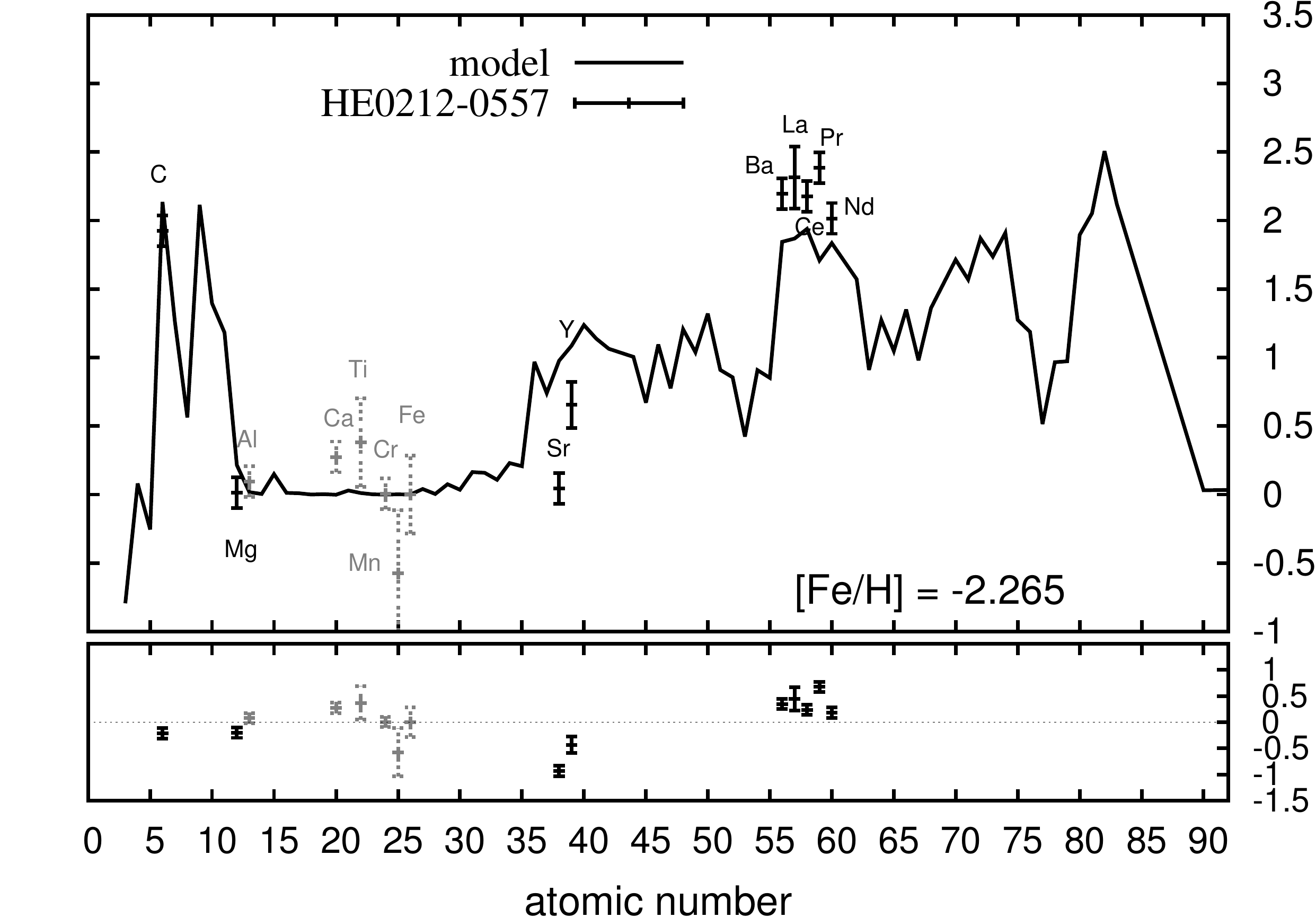}
\caption{Abundances in stars CS$22948-027$ (left panel) and HE$0212-0557$ (right panel). \newline{\it Solid line:} our best fit. {\it Points:} observations. In black we show the elements used to determine the best fit, in grey the elements not affected by AGB nulceosynthesis.}
\label{fig:fits}
\end{figure}

For about $60\%$ of the stars in our sample we found a model that reproduces at the same time the abundances of all the elements involved in AGB nucleosynthesis. An example of a good match between model and observations is shown in the left panel of Fig. \ref{fig:fits}, where the black points are the elements affected by AGB nucleosynthesis, the grey points are the other elements and the solid line is our model. 
For the stars in the other third of our sample we are not able to determine a model that reproduces all the elements simultaneously. In many of these stars, the abundance of light-$s$ elements is low, while heavy-$s$ are strongly enhanced (Fig. \ref{fig:fits}, right). This condition is hard to reproduce in our models, because in the $s$-process the star needs to populate the light-$s$ peak first before it produces heavy-$s$ elements or Pb. One interpretation of this discrepancy is that our parametrization of the PMZ cannot fully take into account all the details of the $s$-process in AGB stars. Further work is required to investigate if a time dependence of $\Mpmz$, or a non-linear mass profile of the H abundance in the PMZ, could lead to a better match between models and observations in all the CEMP-$s$ stars of our sample.

In summary: we implemented the results of the latest models of AGB nucleosynthesis in our binary evolution model to study the observed abundances in a sample of CEMP-$s$ stars. With our models we match the abundance of the elements produced by AGB nucleosynthesis in approximately $60\%$ of the stars of our sample.



\begin{thebibliography}{99}
\bibitem[2013]{Abate2013}
{Abate}, C., {Pols}, O.~R., {Izzard}, R.~G., {Mohamed}, S.~S., \& {de Mink},
  S.~E. 2013, A\&A, 552, A26

\bibitem[{\it in prep.}]{Abate2014}
{Abate}, C., {Pols}, O.~R., {Izzard}, R.~G., \& {Karakas}, A.~I. {\it in prep.}

\bibitem[2007]{Aoki2007}
{Aoki}, W., {Beers}, T.~C., {Christlieb}, N., \etal\ 2007, ApJ, 655, 492

%
%
%
\bibitem[1999]{Busso1999}
{Busso}, M., {Gallino}, R., \& {Wasserburg}, G.~J. 1999, ARA\&A, 37, 239
%
\bibitem[1955]{Cameron1955}{Cameron}, A.~G.~W. 1955, ApJ, 121, 144
%
%
%
\bibitem[2006]{Frebel2006}
{Frebel}, A., {Christlieb}, N., {Norris}, J.~E., \etal\ 2006, ApJ, 652, 1585
%
%
%
%
%
\bibitem[2005]{Herwig2005}{Herwig}, F. 2005, ARA\&A, 43, 435
%
%
\bibitem[2004]{Izzard2004}
{Izzard}, R.~G., {Tout}, C.~A., {Karakas}, A.~I., \& {Pols}, O.~R. 2004,
  MNRAS, 350, 407
%
\bibitem[2006]{Izzard2006}
{Izzard}, R.~G., {Dray}, L.~M., {Karakas}, A.~I., {Lugaro}, M., \& {Tout},
  C.~A. 2006, A\&A, 460, 565
%
\bibitem[2009]{Izzard2009}
{Izzard}, R.~G., {Glebbeek}, E., {Stancliffe}, R.~J., \& {Pols}, O.~R. 2009,
  A\&A, 508, 1359
%
%
\bibitem[2010]{Karakas2010}
{Karakas}, A.~I. 2010, MNRAS, 403, 1413
%
%
%
%
%
\bibitem[2006]{Lucatello2006}
{Lucatello}, S., {Beers}, T.~C., {Christlieb}, N., \etal\ 2006, ApJl, 652,
  L37
%
\bibitem[2005]{Lucatello2005b}
{Lucatello}, S., {Tsangarides}, S., {Beers}, T.~C., \etal\ 2005, ApJ, 625,
  825
%
\bibitem[2012]{Lugaro2012}
{Lugaro}, M., {Doherty}, C.~L., {Karakas}, A.~I., \etal\ 2012, Meteoritics
  and Planetary Science, 47, 1998
%
%
\bibitem[2005]{Marsteller2005}
{Marsteller}, B., {Beers}, T.~C., {Rossi}, S., \etal\ 2005, Nuclear Physics
  A, 758, 312
%
%
%
%
%
%
%
\bibitem[1995]{Straniero1995}
{Straniero}, O., {Gallino}, R., {Busso}, M., \etal\ 1995, ApJl, 440, L85
%
\bibitem[2008]{Suda2008}
{Suda}, T., {Katsuta}, Y., {Yamada}, S., \etal\ 2008, PASJ, 60, 1159
%
\bibitem[2011]{Suda2011}
{Suda}, T., {Yamada}, S., {Katsuta}, Y., \etal\ 2011, MNRAS, 412, 843
%
%
%
%

\end{thebibliography}
\end{document}